\documentclass[12pt]{article}
\usepackage{float}
\usepackage{epsfig,feynmf}

\makeatletter
\@addtoreset{equation}{section}

\renewcommand{\thefootnote}{\fnsymbol{footnote}}
\makeatother

\begin{document}
%%%%%%  user's commands  %%%%%%%%%%%%%%%%%%%%%%%%%%%%%%%%%%%%%%%%%%%
\newcommand{\p}[1]{(\ref{#1})}
\newcommand {\beq}{\begin{eqnarray}}
\newcommand {\eeq}{\end{eqnarray}}
\newcommand {\non}{\nonumber\\}
\newcommand {\eq}[1]{\label {eq.#1}}
\newcommand {\defeq}{\stackrel{\rm def}{=}}
\newcommand {\gto}{\stackrel{g}{\to}}
\newcommand {\hto}{\stackrel{h}{\to}}
\newcommand {\1}[1]{\frac{1}{#1}}
\newcommand {\2}[1]{\frac{i}{#1}}
\newcommand {\thb}{\bar{\theta}}
\newcommand {\ps}{\psi}
\newcommand {\psb}{\bar{\psi}}
\newcommand {\ph}{\varphi}
\newcommand {\phs}[1]{\varphi^{*#1}}
\newcommand {\sig}{\sigma}
\newcommand {\sigb}{\bar{\sigma}}
\newcommand {\Ph}{\Phi}
\newcommand {\Phd}{\Phi^{\dagger}}
\newcommand {\Sig}{\Sigma}
\newcommand {\Phm}{{\mit\Phi}}
\newcommand {\eps}{\varepsilon}
\newcommand {\del}{\partial}
\newcommand {\dagg}{^{\dagger}}
\newcommand {\pri}{^{\prime}}
\newcommand {\prip}{^{\prime\prime}}
\newcommand {\pripp}{^{\prime\prime\prime}}
\newcommand {\prippp}{^{\prime\prime\prime\prime}}
\newcommand {\pripppp}{^{\prime\prime\prime\prime\prime}}
\newcommand {\delb}{\bar{\partial}}
\newcommand {\zb}{\bar{z}}
\newcommand {\mub}{\bar{\mu}}
\newcommand {\nub}{\bar{\nu}}
\newcommand {\lam}{\lambda}
\newcommand {\lamb}{\bar{\lambda}}
\newcommand {\kap}{\kappa}
\newcommand {\kapb}{\bar{\kappa}}
\newcommand {\xib}{\bar{\xi}}
\newcommand {\ep}{\epsilon}
\newcommand {\epb}{\bar{\epsilon}}
\newcommand {\Ga}{\Gamma}
\newcommand {\rhob}{\bar{\rho}}
\newcommand {\etab}{\bar{\eta}}
\newcommand {\chib}{\bar{\chi}}
\newcommand {\tht}{\tilde{\th}}
\newcommand {\zbasis}[1]{\del/\del z^{#1}}
\newcommand {\zbbasis}[1]{\del/\del \bar{z}^{#1}}
\newcommand {\vecv}{\vec{v}^{\, \prime}}
\newcommand {\vecvd}{\vec{v}^{\, \prime \dagger}}
\newcommand {\vecvs}{\vec{v}^{\, \prime *}}
\newcommand {\alpht}{\tilde{\alpha}}
\newcommand {\xipd}{\xi^{\prime\dagger}}
\newcommand {\pris}{^{\prime *}}
\newcommand {\prid}{^{\prime \dagger}}
\newcommand {\Jto}{\stackrel{J}{\to}}
\newcommand {\vprid}{v^{\prime 2}}
\newcommand {\vpriq}{v^{\prime 4}}
\newcommand {\vt}{\tilde{v}}
\newcommand {\vecvt}{\vec{\tilde{v}}}
\newcommand {\vecpht}{\vec{\tilde{\phi}}}
\newcommand {\pht}{\tilde{\phi}}
\newcommand {\goto}{\stackrel{g_0}{\to}}
\newcommand {\tr}{{\rm tr}\,}
\newcommand {\GC}{G^{\bf C}}
\newcommand {\HC}{H^{\bf C}}
\newcommand{\vs}[1]{\vspace{#1 mm}}
\newcommand{\hs}[1]{\hspace{#1 mm}}
\newcommand{\al}{\alpha}
\newcommand{\be}{\beta}
\newcommand{\Lam}{\Lambda}
\newcommand{\kahler}{K\"ahler }
\newcommand{\con}[1]{{\Gamma^{#1}}}
\thispagestyle{empty}

\begin{flushright}
{\tt hep-ph/0409273} \\
{\tt KEK-TH-981}
\end{flushright}
\vspace{2mm}

\begin{center}
{\Large
{\bf Top Spin Correlations 
 in Theories with Large Extra-Dimensions 
 at the Large Hadron Collider
}
}
\\[8mm]
\vspace{3mm}
\normalsize
{\large \bf
  Masato~Arai~$^{a}$}
\footnote{\it
arai@fzu.cz}
,
{\large \bf
  Nobuchika~Okada~$^{b}$}
\footnote{\it
okadan@post.kek.jp}
,
{\large \bf Karel Smolek~$^{c}$}
\footnote{\it
karel.smolek@utef.cvut.cz
} \\
and \\
 {\large \bf
Vladislav \v{S}im\'ak~$^{a,d}$}
\footnote{\it
simak@fzu.cz
}

\vskip 1.5em
{$^{a}$ \it Institute of Physics, AS CR,
        182 21, Prague 8, Czech Republic \\
 $^{b}$ Theory Division, KEK, Tsukuba, 
        Ibaraki, 305-0801, Japan \\
 $^{c}$ \it Institute of Experimental and Applied Physics, \\
        Czech Technical University in Prague, 
        128 00, Prague 2, Czech Republic \\
 $^{d}$ \it Faculty of Nuclear Sciences and Physical Engineering,\\
        Czech Technical University in Prague, 
        115 19, Prague 1, Czech Republic 
}
\vspace{3mm}\\
{\bf Abstract}\\[5mm]
{\parbox{13.5cm}{
In theories with large extra dimensions, 
 we study the top spin correlations 
 at the Large Hadron Collider. 
The $s$-channel process mediated 
 by graviton Kaluza-Klein modes 
 contributes to the top-antitop pair production 
 in addition to the Standard Model processes, 
 and affects the resultant top spin correlations. 
We calculated the full density matrix for the top-antitop pair
 production. 
With the fundamental scale of the extra dimensional theory 
 below $2$ TeV, 
 we find a sizable deviation of the top spin correlations 
 from the Standard Model one.
}}
\end{center}
\vfill
\newpage
\setcounter{page}{1}
\setcounter{footnote}{0}
\renewcommand{\thefootnote}{\arabic{footnote}}
%%%%%%%%%%%%%%%%%%%%%%%%%%%%%%%
%
% Introduction
%
%%%%%%%%%%%%%%%%%%%%%%%%%%%%%%%
\section{Introduction}
Recently, the possibility that our space has more 
 than three spatial dimensions
 has been vigorously studied \cite{LED,RS}.
Typical setup is that the standard model (SM) fields reside 
 on $3+1$-dimensional manifold called ``3-brane'' embedded
 in higher dimensional space-time 
 while only graviton can propagate into the whole space-time dimensions.
This setup opens up a new way to solve
 the gauge hierarchy problem, 
 namely, a huge
 hierarchy between the electroweak scale and the
 4-dimensional Planck scale, 
 as addressed by Arkani-Hamed, Dimopoulos and Dvali (ADD)
 \cite{LED}.
In their model,
 $4$-dimensional Planck scale $M_{pl}$ can be expressed 
 by the fundamental scale of $4+n$ dimensions $M_D$ 
 and the common radius of the compactified $n$ extra dimensions $R$ 
 such as $M_{pl}=M_D(M_D R)^{n/2}$. 
If the compactification radius is large enough 
 (for example, $R\sim 0.1$ mm for $n=2$), 
 $M_D$ can be ${\cal O}(1 {\rm TeV})$ and 
 thus the gauge hierarchy problem can be solved.

Effects of extra dimensions are recast 
 in four dimensional effective theory 
 valid below $M_D$ \cite{Rattazzi,Han}. 
In the effective theory, the graviton propagating 
 in the whole $4+n$ dimensions 
 can be expressed as an infinite tower of Kaluza-Klein (KK) modes 
 which contain spin-2 (KK-graviton), spin-1 (gravi-vector)
 and spin-0 (gravi-scalar) excitations.
The KK gravitons and the gravi-scalars couple to the 
 energy-momentum tensor of the SM fields and its trace, respectively, 
 while the gravi-vector has no interaction with them. 
Since the trace of the energy-momentum tensor
 is vanishing for massless fields, we can ignore 
 the gravi-scalar interaction at high energy processes.
Although each vertex between the SM fields and the KK gravitons 
 are suppressed by $M_{pl}$,
 the effective coupling is enhanced 
 due to contributions of a large number of KK gravitons 
 so as to be suppressed by $M_D={\cal O}(1 {\rm TeV})$. 
Therefore, new phenomena induced by the KK gravitons, 
 for example, 
 the direct KK graviton emission processes  
 and the virtual KK graviton exchange processes,
 can be expected at high energy collisions \cite{Cheung}. 
In particular, 
 the virtual KK graviton exchange process is interesting, 
 since it can give rise to characteristic spin configurations 
 and angular distributions for outgoing particles, 
 which reflect the spin-2 nature of the intermediate KK gravitons. 

The top-antitop quark pair is a good candidate 
 to study its spin correlations, 
 since the top quark, with mass in the range of $175$ GeV \cite{Abe}, 
 decays electroweakly before hadronizing \cite{Bigi},
 and thus the possible polarization of the top-antitop quark pair 
 is transferred to its decay products. 
There are thus significant angular correlations 
 among the top quark spin axis and 
 the direction of motion of the decay products.

For the hadronic top-antitop pair production process 
 through the quantum chromodynamics (QCD) interaction, 
 the spin correlations have been extensively studied 
 \cite{Stelzer, Mahlon-Parke, Bernreuther2}. 
It is found that there is a spin asymmetry between 
 the produced top-antitop pairs, namely, 
 the number of produced top-antitop quark pairs 
 with both spin up or spin down 
 is different from the number of pairs 
 with the opposite spin combinations. 
If the top quark is coupled to a new physics beyond the SM, 
 the top-antitop spin correlations could be altered. 
Therefore, the top-antitop correlations are 
 useful information   
 to test the SM and also a possible new physics 
 at hadron colliders. 
Note that the Large Hadron Collider (LHC) 
 has a big advantage to study them, 
 since it will produce almost 10 millions top quarks per year. 

In the ADD model, there exists 
 a new top-antitop pair production process 
 through the virtual KK graviton exchange in the {\it s}-channel. 
The effect of the virtual KK graviton 
 exchange process has been studied 
 for the total cross section of the top-antitop pair production 
 process at hadron collider \cite{Mathews} 
 and for polarized amplitudes at $e^+e^-$ 
 and $\gamma\gamma$ colliders \cite{Lee1,Lee2}. 
The purpose of this paper is 
 to examine the effect of the virtual KK graviton exchange process 
 on the spin correlation of the top-antitop pairs 
 produced at the LHC. 
With the fundamental scale of the ADD model 
 below $2$ TeV, 
 we find a sizable deviation of the spin asymmetry of 
 produced top-antitop quark pairs from the SM one.
Furthermore, since, in practice, information of the spins 
 of the produced top-antitop quarks are measured 
 through the angular correlations among the top spin axis and
 the direction of motion of the decay products of the top quark, 
 we study them also.

This paper is organized as follows. 
In Sec. 2, we briefly review the top spin correlations.
In Sec. 3, we examine the invariant amplitudes 
 for the polarized top-antitop quark pair production 
 processes, 
 $q\bar{q}\rightarrow t \bar{t}$ and $gg\rightarrow t\bar{t}$, 
 including the process mediated by the virtual KK gravitons 
 in the {\it s}-channel. 
We perform numerical analysis in Sec. 4. 
Sec. 5 is devoted to conclusions. 
In Appendix, full representation of the density 
 matrix (squared invariant amplitude) for the
 top-antitop pair production is given.

%%%%%%%%%%%%%%%%%%%%%%%%%%%%%%%
%
% Spin correlation
%
%%%%%%%%%%%%%%%%%%%%%%%%%%%%%%%
\section{Spin correlation}
At hadron collider, the top-antitop quark pair is produced 
 through the processes of quark-antiquark pair annihilation 
 and gluon fusion: 
\begin{eqnarray}
 &i \rightarrow t+\bar{t}, \,\,\, i=q\bar{q}\,,gg\,.& \label{top1}
\end{eqnarray}
The former is the dominant process at the Tevatron, 
 while the latter is dominant at the LHC. 
The produced top-antitop pairs decay  
 before hadronization takes place. 
The main decay modes in the SM 
 involve leptonic and hadronic modes: 
\begin{eqnarray}
 &t\rightarrow bW^+ \rightarrow bl^+\nu_l\,,bu\bar{d}\,,bc\bar{s},&
 \label{decay}
\end{eqnarray}
where $l=e,\mu,\tau$.
The differential decay rates to a decay product $f=b,l^+, \nu_l,$ etc.  
 in the top quark rest frame can be parameterized as 
\begin{eqnarray}
{1 \over \Gamma}{d \Gamma \over d \cos \theta_f}=
  {1 \over 2}(1 + \kappa_f \cos \theta_f ), 
 \label{decay1}
\end{eqnarray}
where $\Gamma$ is the partial decay width of 
 the respective decay channel,  
 $\theta_f$ is the angle between the chosen top spin axis 
 and the direction of motion of the decay product $f$,
 and  
 the coefficient $\kappa_f$ is the top-spin analyzing power 
 of a particle $f$. 
The SM values of $\kappa_f$ at tree level 
 have been computed \cite{Jezabek}, 
 for instance, in the semi-leptonic decay, 
 $\kappa_{l^+}= +1$ for the charged lepton, 
 $\kappa_b = -0.41$ for the $b$-quark 
 and $\kappa_{\nu_l}=-0.31$ for the $\nu_l$, respectively. 
In hadronic decay modes, the role of the charged lepton 
 is replaced by the $d$ or $s$ quark. 

The total matrix element squared 
 for the top-antitop pair production \p{top1} 
 and their decay channels \p{decay} is given by 
\begin{eqnarray}
|{\cal M}|^2 \propto {\rm Tr}[\rho R^i \bar{\rho}]
 =\rho_{\alpha^\prime\alpha}R^i_{\alpha\beta,\alpha^\prime\beta^\prime}
  \bar{\rho}_{\beta^\prime\beta}. \label{comp}
\end{eqnarray}
Here the subscripts denote the top and antitop spin indices, 
 and $R^i$ denotes the density matrix 
 corresponding to the production of the on-shell top-antitop quark pair 
 through the process $i$ in \p{top1}:
\begin{eqnarray}
 R_{\alpha\beta,\alpha^\prime\beta^\prime}^i
 =\sum_{{\rm initial~spin}}{\cal M}(i\rightarrow t_\alpha\bar{t}_\beta)
  {\cal M}^*(i\rightarrow t_{\alpha^\prime}\bar{t}_{\beta^\prime}),
\end{eqnarray}
where ${\cal M}(i\rightarrow t_\alpha\bar{t}_\beta)$ is 
 the amplitude for the top-antitop pair production. 
The matrices $\rho$ and $\bar{\rho}$ are the density matrices 
 corresponding to the decays of polarized top and antitop quarks 
 into some final states at the top and antitop rest frame, respectively.
In the leptonic decay modes, 
 the matrices $\rho$, which leads to \p{decay1},
 can be obtained as (see, for instance, \cite{Bernreuther})
\begin{eqnarray}
  \rho_{\alpha^\prime\alpha}
  = {\cal M}(t_\alpha \rightarrow bl^+\nu_l)
    {\cal M}^*(t_{\alpha^\prime} \rightarrow bl^+\nu_l)  
  = {\Gamma \over 2}(1 + \kappa_f {\vec{\sigma}} \cdot 
       \vec{q}_f)_{\alpha^\prime\alpha},  
 \label{rho1}
\end{eqnarray}
where $q_f$ is the unit vector of the direction of motion 
 of the decay product $f$. 
The density matrix for the polarized antitop quark 
 is obtained by replacing $\kappa_f \rightarrow -\kappa_f$ in \p{rho1}.

It is clear that the best way to analyze 
 the top-antitop spin correlations 
 is to see the angular correlations 
 of two charged leptons $l^+l^-$ 
 produced by the top-antitop quark leptonic decays. 
In the following, we consider only the leptonic decay channels. 
Using \p{comp}-\p{rho1} and integrating over 
 the azimuthal angles of the charged leptons, 
 we obtain the following double distribution
 \cite{Stelzer, Mahlon-Parke, Bernreuther2}
\begin{eqnarray}
 {1 \over \sigma}
 {d^2 \sigma \over d \cos\theta_{l^+} d \cos\theta_{l^-}}
 = {1-{\cal A} \kappa_{l^+}\kappa_{l^-}
    \cos\theta_{l^+} \cos\theta_{l^-} \over 4 }, 
 \label{double}
\end{eqnarray}
with $ \kappa_{l^+}=\kappa_{l^-}=1$. 
Here $\sigma$ denotes the cross section 
 for the process of the leptonic decay modes, 
 and $\theta_{l^+} (\theta_{l^-})$ denotes the angle 
 between the top (antitop) spin axis and  
 the direction of motion of the antilepton (lepton) 
 in the top (antitop) rest frame. 
The coefficient ${\cal A}$ denotes 
 the spin asymmetry between 
 the produced top-antitop pairs 
 with like and unlike spin pairs defined as 
\begin{eqnarray}
 {\cal A}={\sigma(t_\uparrow\bar{t}_\uparrow)
          +\sigma(t_\downarrow\bar{t}_\downarrow) 
          -\sigma(t_\uparrow\bar{t}_\downarrow)
          -\sigma(t_\downarrow\bar{t}_\uparrow) 
         \over
           \sigma(t_\uparrow\bar{t}_\uparrow)
          +\sigma(t_\downarrow\bar{t}_\downarrow)
          +\sigma(t_\uparrow\bar{t}_\downarrow)
          +\sigma(t_\downarrow\bar{t}_\uparrow)} , 
\label{asym}
\end{eqnarray}
where $\sigma(t_\alpha\bar{t}_{\dot{\alpha}})$ is 
 the cross section of the top-antitop pair production 
 at parton level
 with denoted spin indices.

In the SM, at the lowest order of $\alpha_s$,
 the spin asymmetry is found to be ${\cal A}=+0.302$ 
 for the LHC. 
\footnote{
The parton distribution function set of CTEQ5M1 \cite{CTEQ} 
 has been used in our calculations. 
The resultant spin asymmetry somewhat depends 
 on the parton distribution functions used. } 
Since in the ADD model 
 there is a new contribution 
 to the top-antitop pair production process \p{top1} 
 through the virtual KK graviton exchange in the {\it s}-channel, 
 the spin asymmetry \p{asym} can be altered from the SM one. 
In the next section, we calculate the squared amplitudes 
 for the top-antitop pair production 
 including the virtual KK graviton mediated process. 

%%%%%%%%%%%%%%%%%%%%%%%%%%%%%%%
%
% Amplitude
%
%%%%%%%%%%%%%%%%%%%%%%%%%%%%%%%
\section{Scattering Amplitude} 
In the 4-dimensional effective theory of the ADD model, 
 the interaction Lagrangian 
 between the $n$-th KK mode of graviton $G^{(\vec{n})}$ 
 and the SM fields is given by 
\begin{eqnarray}
 {\cal L}_{int} = -{1 \over M_P}\sum_{\vec{n}}
                    G_{\mu\nu}^{(\vec{n})}T^{\mu\nu}, \label{int}
\end{eqnarray}
where $M_P=M_{pl}/\sqrt{8 \pi}$ is 
 the reduced 4-dimensional Planck scale, 
 and $T^{\mu\nu}$ is energy momentum tensor of the SM fields.
Thus, there exists the top-antitop pair production process 
 $\lambda(k_1)+\bar{\lambda}(k_2)
 \rightarrow t(k_3) + \bar{t}(k_4)$ 
 (where $\lambda$ denotes quark or gluon) 
 through the virtual KK graviton exchange in the {\it s}-channel. 
In the 4-dimensional effective theory 
 with energies smaller than 
 the 4+n dimensional Planck scale $M_D$, 
 the amplitude for the KK graviton exchange process 
 would be described as \cite{Rattazzi} 
\begin{eqnarray}
{\cal M}_G =
 \frac{4 \pi \lambda}{M_D^4}  
 T^{\mu \nu} (k_1 , k_2) T_{\mu \nu} (k_3 , k_4). 
 \label{amplitude2} 
\end{eqnarray}
All the ambiguity such as the number of extra dimensions 
 and the regularization procedure for  
 the contributions from the infinite number of KK gravitons 
 are encoded by an order one parameter $\lambda$. 
Hereafter we use the effective scattering amplitude \p{amplitude2} 
 for the top-antitop pair production process 
 and consider the two cases $\lambda=\pm 1$. 
Using \p{amplitude2}, 
 the amplitudes of the top-antitop pair production process 
 through the virtual KK graviton exchanges 
 can be explicitly described as (color and flavor 
 indices are suppressed)
\begin{eqnarray}
 {\cal M}_G(q\bar{q} \rightarrow t\bar{t}) 
      &=& f_G
          {\Big \{}[\bar{\psi}(k_2)\gamma^\mu\psi(k_1)]
                    [\bar{\psi}(k_3)\gamma_\mu\psi(k_4)]
                    (k_1 - k_2)^\nu (k_3 - k_4)_\nu \non
      & & ~~+
          [\bar{\psi}(k_3) \gamma^\mu (k_1-k_2)_\mu \psi(k_4)]
          [\bar{\psi}(k_2) \gamma^\nu (k_3-k_4)_\nu \psi(k_1)]
          {\Big \}}, \label{qamp} \\
 {\cal M}_G(gg \rightarrow t\bar{t})
      &=& 4 f_G
          {\Big \{}-[\bar{\psi(k_3)}\gamma^\mu (k_3-k_4)_\mu \psi(k_4)]  
                    (k_1\cdot k_2)[\epsilon(k_1)\cdot \epsilon(k_2)] \non
      & &+ [\bar{\psi}(k_3)\gamma^\mu k_{2\mu} \psi(k_4)][k_1\cdot (k_3-k_4)]
           [\epsilon(k_1)\cdot \epsilon(k_2)] \non 
      & &+ [\bar{\psi}(k_3)\gamma^\mu k_{1\mu} \psi(k_4)][k_2\cdot (k_3-k_4)]
           [\epsilon(k_1)\cdot \epsilon(k_2)] \non
      & &+ [\bar{\psi}(k_3)\gamma^\mu \epsilon_{\mu}(k_2) \psi(k_4)]
           (k_1\cdot k_2)
           [(k_3 - k_4) \cdot \epsilon(k_1)] \non
      & &+ [\bar{\psi}(k_3)\gamma^\mu \epsilon_{\mu}(k_1) \psi(k_4)]
           (k_1\cdot k_2)
           [(k_3 - k_4) \cdot \epsilon(k_2)] {\Big \}}, \label{gamp}
\end{eqnarray}
where $f_G \equiv \pi \lambda /2 M_D^4$, 
 and $\epsilon(k_i)~(i=1,2)$ are the polarization vectors 
 of the initial gluons.
   
In the center of mass frame, 
 one can straightforwardly calculate 
 the density matrix $R_{\alpha\beta,\alpha^\prime\beta^\prime}^i$
 including both the SM (QCD) and the virtual KK graviton 
 contributions. 
In the following, 
 we give the diagonal parts of the density matrix 
 which are relevant for the estimation of \p{double} and \p{asym}.
Full representation of the density matrix is given in Appendix. 
In our calculation, 
 we have chosen the helicity spin basis 
 useful for the energy range at the LHC, 
 while the off-diagonal basis is suitable 
 for Tevatron \cite{parke-shadmi}. 
The results are in the following. 
For the process $q\bar{q} \rightarrow t \bar{t}$, we find
\begin{eqnarray}
 |{\cal M}(q\bar{q}\rightarrow t_\uparrow\bar{t}_\uparrow)|^2
  &=& |{\cal M}
 (q\bar{q}\rightarrow {t_\downarrow\bar{t}_\downarrow})|^2 \non
  &=& {g_s^4 \over 9}(1-\beta^2)\sin^2\theta
 +{f_G^2 s^4\beta^2 \over 2}(1-\beta^2)\sin^22\theta, 
  \label{qq1} \\
 |{\cal M}(q\bar{q}\rightarrow t_\uparrow\bar{t}_\downarrow)|^2
 &=& |{\cal M}(q\bar{q}\rightarrow {t_\downarrow\bar{t}_\uparrow})|^2 \non
  &=& {g_s^4 \over 9}(1+\cos^2\theta)
 +{f_G^2 s^4\beta^2 \over 2}
    (\cos^22\theta+\cos^2\theta),
  \label{qq2}
\end{eqnarray}
where $\theta$ is the scattering angle 
 between the incoming $q$ and outgoing $t$, 
 $g_s$ is the QCD coupling constant, 
 $\beta=\sqrt{1-4m^2_t/s}$, and $m_t$ is top quark mass. 
For the $gg$ initial state, we obtain 
\begin{eqnarray} 
 &&|{\cal M}(gg \rightarrow t_\uparrow\bar{t}_\uparrow)|^2
    = |{\cal M}(gg \rightarrow {t_\downarrow\bar{t}_\downarrow})|^2 \non
 &&~~~~
    ={g_s^4 \beta^2 \over 96}{\cal Y}(\beta,\cos\theta)(1-\beta^2)
     (1+\beta^2+\beta^2\sin^4\theta)
     +{\cal Z}(\beta,\theta,s)
     f_G s^2 \beta^2(1-\beta^2)\sin^4\theta, \label{gg1} \\
 &&|{\cal M}(gg\rightarrow t_\uparrow\bar{t}_\downarrow)|^2
   = |{\cal M}(q\bar{q}\rightarrow {t_\downarrow\bar{t}_\uparrow})|^2 \non
 &&~~~~
     ={g_s^4 \beta^2 \over 96}{\cal Y}(\beta,\cos\theta)\sin^2\theta(1+\cos^2\theta)
      +{\cal Z}(\beta,\theta,s)
      f_Gs^2 \beta^2\sin^2\theta(1+\cos^2\theta). \label{gg2}
\end{eqnarray}
Here $ {\cal Y}(\beta,\theta,s)$ and $ {\cal Z}(\beta,\theta,s)$ 
 are defined by
\begin{eqnarray}
 {\cal Y}(\beta,\cos\theta)=
  {7+9\beta^2\cos^2 \theta \over (1-\beta^2\cos^2\theta)^2},~~
 {\cal Z}(\beta,\theta,s)={g_s^2 \over 4(1-\beta^2\cos^2\theta)}
       +{3 \over 4}f_G s^2, \label{funcs}
\end{eqnarray}
respectively.
Note that there is no interference between the SM diagram 
 and the virtual KK graviton exchange diagram 
 in the $q\bar{q}$ initiated process, 
 while there is the non-vanishing interference 
 in the $gg$ initiated process. 

With the squared amplitudes in \p{qq1}, \p{qq2}, \p{gg1} and \p{gg2},
 the integrated top-antitop quark pair production cross section 
 can be obtained by using the formula
\begin{eqnarray}
 \sigma_{tot}(pp \rightarrow t_\alpha\bar{t}_{\dot{\alpha}})=
  \sum_{a,b} \int dx_1 \int dx_2 \int d \cos \theta 
   f_a(x_1,Q^2)f_b(x_2,Q^2)\nonumber \\
   \times {d \sigma(a(x_1E_{CMS})b(x_2E_{CMS})
   \rightarrow t_\alpha\bar{t}_{\dot{\alpha}}) \over d \cos \theta} 
   \label{total}
\end{eqnarray} 
where $f_a$ is the parton distribution function 
 for a parton $a$, $E_{CMS}$ is a center-of-mass energy of a proton,
 and $Q$ is the virtual momentum transfer.

\begin{figure}[H]
\begin{center}
  \epsfxsize=7.5cm
  \epsfbox{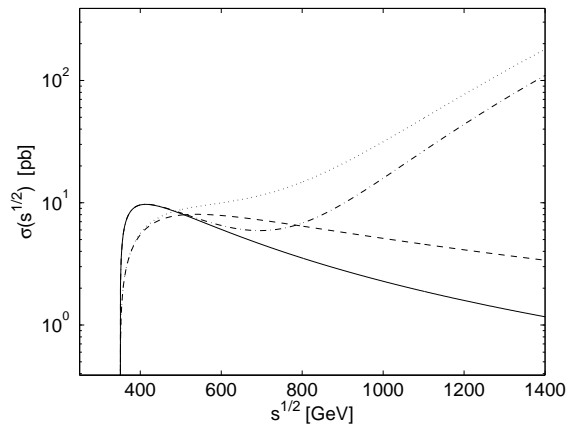}
\caption{The dependence of the total cross section of the 
 top-antitop quark pair production 
 on the center-of-mass energy of colliding partons. 
The solid line and the dashed line correspond to 
 the SM prediction for the $q\bar{q}\rightarrow t\bar{t}$ 
 and the $gg\rightarrow t\bar{t}$ channels, respectively. 
The dash-dotted line and the dotted line correspond to 
 the predictions of the ADD model 
 for the same channels 
 with $\lambda=1$ and $M_D=1$ TeV.}
  \label{fig_sigma_s}
\end{center}
\end{figure}

As can be seen from the formulas of the squared amplitudes, 
 the cross sections through the virtual KK graviton exchange
 process grow according to a power of 
 the collider center-of-mass energy, 
 and thus the unitarity will be violated at 
 high energies.\footnote{
 See, for example, \cite{Eboli} for discussions on the unitarity 
 in theories with large extra dimensions.} 
This behavior is shown in Fig.~\ref{fig_sigma_s}, 
 where 
 the total cross sections of the top-antitop pair production 
 through $q\bar{q}\rightarrow t\bar{t}$ and 
 $gg\rightarrow t\bar{t}$ at the parton level, respectively, 
 are depicted as a function of 
 parton center-of-mass energy $\sqrt{s}$ 
 (or, equivalently, invariant mass 
 of the produced top-antitop pair $M_{t\bar{t}}$). 
We can see that 
 the cross section of the ADD model grows rapidly with $\sqrt{s}$, 
 while the SM cross section decreases. 
As mentioned above, 
 we (should) use the formula of (\ref{amplitude2}) 
 at energies lower than $M_D$. 
Therefore, in order to make our analysis conservative, 
 we take into account the contributions 
 from the virtual KK graviton exchange processes 
 only for the center-of-mass energy of colliding partons 
 lower than $M_D$, namely $\sqrt{s}=M_{t\bar{t}} \leq M_D$. 

%%%%%%%%%%%%%%%%%%%%%%%%%%%%%%%%%%%%%%%%%%%%%%%%%%
%
% Numerical results
%
%%%%%%%%%%%%%%%%%%%%%%%%%%%%%%%%%%%%%%%%%%%%%%%%%%
\section{Numerical Results}

Here we present various numerical results 
 and demonstrate interesting properties 
 of measurable quantities in the ADD model. 
In the following analysis, we use the parton distribution functions 
 of Ref. \cite{CTEQ} (CTEQ5M1) and 
 its numerical implementation in PDFLIB \cite{PDFLIB} 
 from the CERN Program Library 
 with the constant scale $Q=m_t=175~{\rm GeV}$, $N_f=5$
 and $\alpha_s(Q)=0.1074$.

\begin{figure}[H]
\begin{center}
  \epsfxsize=7.5cm
  \epsfbox{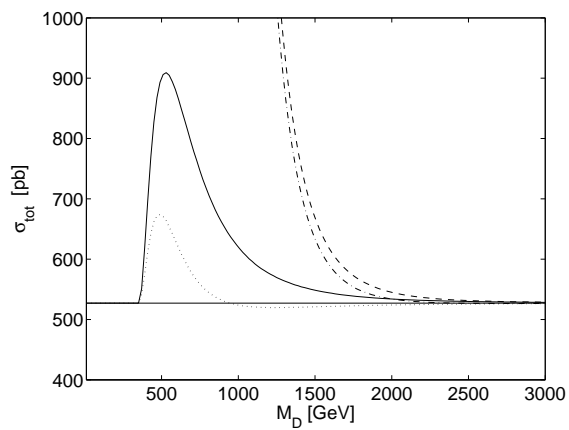}
\end{center}
\caption{Total cross section for the top pair production 
 as a function of the scale $M_D$ 
 for the LHC with center-of-mass energy $14$ TeV.
The horizontal line corresponds to  
 the prediction of the SM, and the
 solid (dotted) line to the ADD model in 
 $\lambda=1$ ($\lambda=-1$) case with 
 the cut. 
The dashed (dash-dotted) line depicts 
 the plot in $\lambda=1$ ($\lambda=-1$) case
 without the cut. }
\label{cross}
\end{figure}

\begin{figure}[H]
\begin{center}
\begin{eqnarray*}
 \begin{array}{ccc}
  \epsfxsize=7.5cm
  \epsfbox{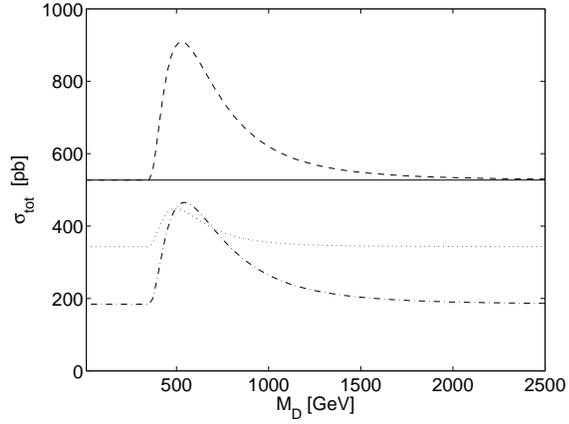}
  & 
  \hspace{3mm}
  & 
  \epsfxsize=7.5cm
  \epsfbox{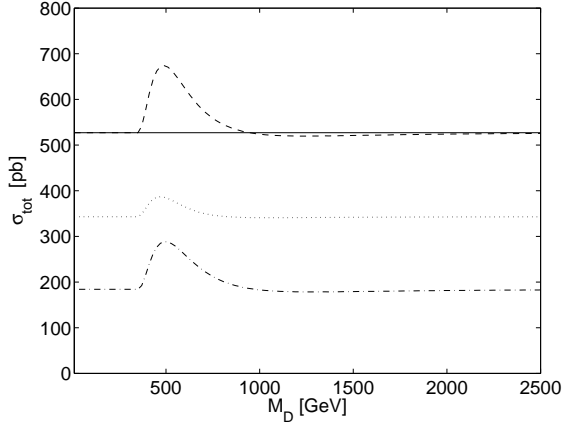} \\
  \mbox{\footnotesize (a)}~\lambda=1
  &
  \hspace{3mm}
  &
  \mbox{\footnotesize (b)}~\lambda=-1
 \end{array}
\end{eqnarray*}
\end{center}
\caption{The breakdown of the total 
 cross section into the like 
 (dotted) and the unlike (dash-dotted) top-antitop  
 spin pairs in the ADD model 
 besides the total cross section of
 the SM (horizontal lines) and the ADD model (dashed lines) 
 results with $M_D=1$ TeV and
 $E_{CMS}=14$ TeV.}
\label{cross2}
\end{figure}

Figs.~\ref{cross} and \ref{cross2} show 
 the total cross sections of the top-antitop quark pair production
 as a function of the scale $M_D$ 
 at the LHC with the center-of-mass energy $14$ TeV. 
In Fig.~\ref{cross}, we present the plots with and without the cut 
 $M_{t\bar{t}}\leq M_D$ for the KK graviton mediated process. 
In both cases,
 the cross sections in the ADD model 
 trace the SM line for large $M_D$.
The total cross section without the cut
 grows rapidly
 as the scale $M_D$ becomes small.
On the other hand, the cross section with 
 the cut traces the SM result for small $M_D$.
It indicates that our analysis is conservative.
In Fig. \ref{cross2}, we show the breakdown of the total 
 cross section into the like 
 ($t_{\uparrow}\bar{t}_{\uparrow}
 +t_{\downarrow}\bar{t}_{\downarrow}$) and 
 the unlike top-antitop 
 ($t_{\uparrow}\bar{t}_{\downarrow}
 +t_{\downarrow}\bar{t}_{\uparrow}$) 
 spin pairs in the ADD model 
 besides the total cross section of
 the SM and the ADD model results.
For the scale $M_D$ below $2$ TeV, 
 we can see sizable deviations from the SM one.

\begin{figure}[H]
\begin{center}
\begin{eqnarray*}
 \begin{array}{ccc}
  \epsfxsize=7.2cm
  \epsfbox{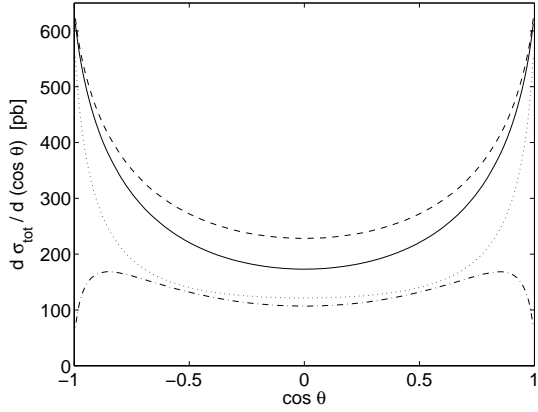}
  & 
  \hspace{3mm}
  & 
  \epsfxsize=7.2cm
  \epsfbox{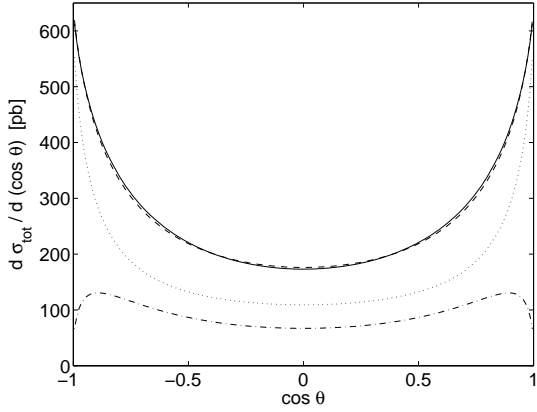} \\
  \mbox{\footnotesize (a)}~\lambda=1
  &
  \hspace{3mm}
  &
  \mbox{\footnotesize (b)}~\lambda=-1
 \end{array}
\end{eqnarray*}
\caption{Differential cross section \p{total_cosTheta} 
 as a function of $\cos\theta$ with $M_D=1$ TeV and $E_{CMS}=14$ TeV.
The solid lines and dashed lines correspond to 
 the results of the SM and the ADD model, respectively. 
The differential cross sections 
 for the like (dotted) and the unlike (dash-dotted) 
 top-antitop spin pair productions in the ADD model 
 are also depicted. }
\label{cross3}
\end{center}
\end{figure}

Differential cross section for the top-antitop pair production 
 given by
\begin{eqnarray}
 \frac{d \sigma_{tot}(pp \rightarrow t\bar{t})}{d\cos\theta}=
  \sum_{a,b} \int d x_1 \int d x_2\, 
   f_a(x_1,Q^2)f_b(x_2,Q^2)
   {d\sigma(t\bar{t}) \over d\cos \theta} 
   \label{total_cosTheta}
\end{eqnarray}
 is shown in Fig.~\ref{cross3} 
 with the scale $M_D=1~{\rm TeV}$ and $E_{CMS}=14$ TeV.
In $\lambda = 1$ case, 
 we can see the large deviation from the SM cross section 
 at large scattering angles.  
On the other hand, the deviation for the SM cross section 
 is small in $\lambda=-1$ case. 
This is because a cancellation occurs 
 between the interference term 
 among the SM and the KK graviton mediated processes 
 and the purely KK graviton mediated process. 

\begin{figure}[H]
\begin{center}
\begin{eqnarray*}
 \begin{array}{ccc}
  \epsfxsize=7.2cm
  \epsfbox{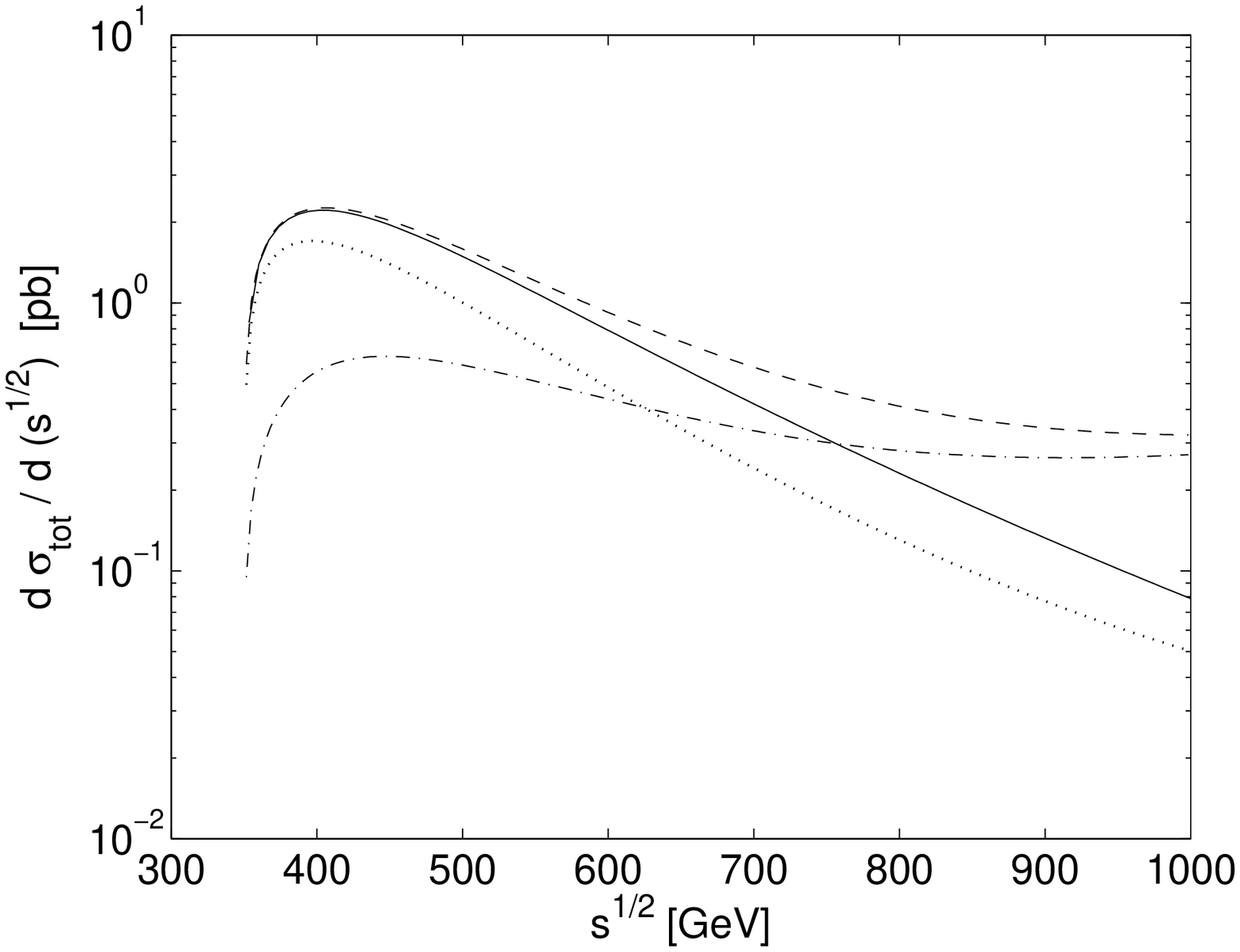}
  & 
  \hspace{3mm}
  & 
  \epsfxsize=7.2cm
  \epsfbox{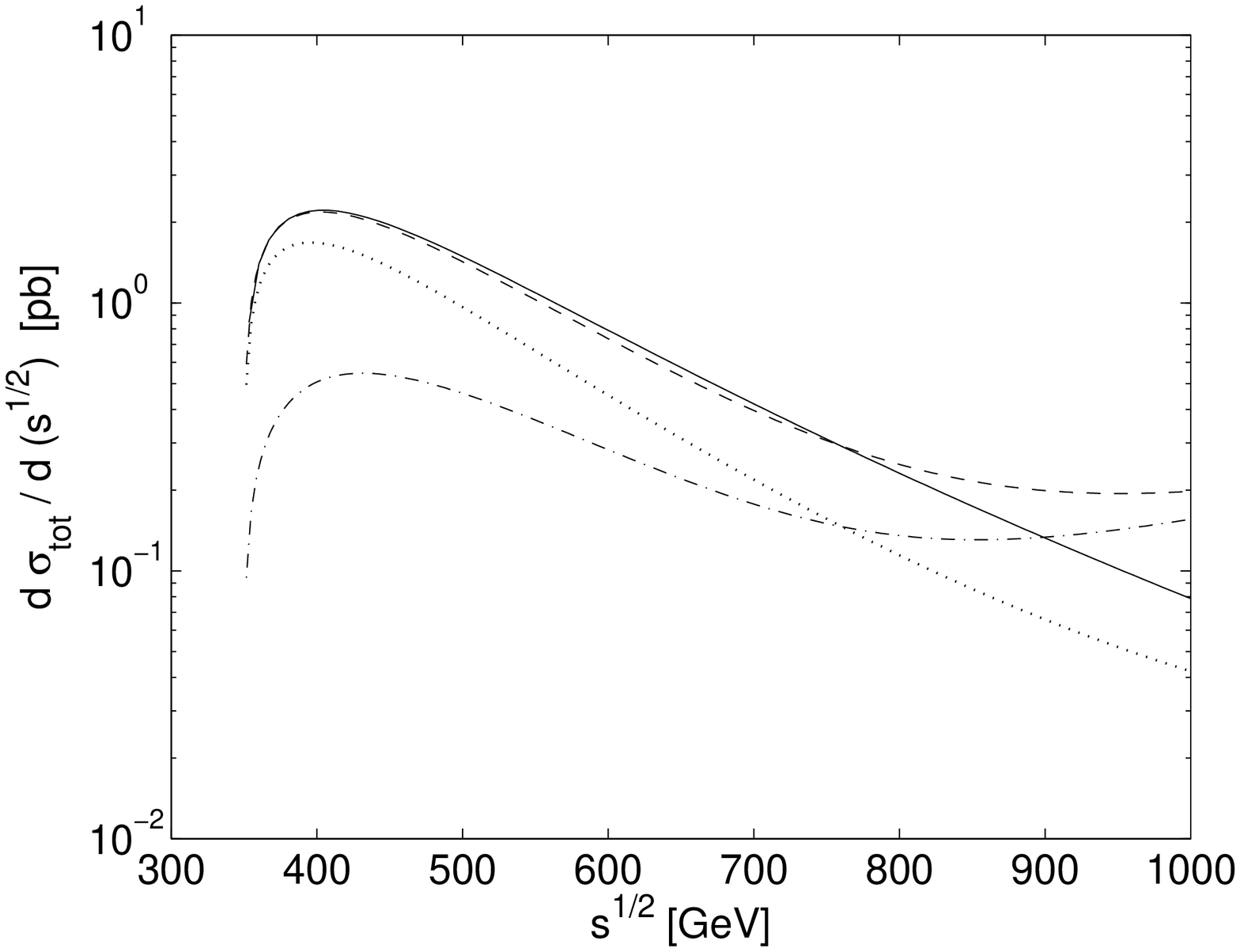} \\
  \mbox{\footnotesize (a)}~\lambda=1
  &
  \hspace{3mm}
  &
  \mbox{\footnotesize (b)}~\lambda=-1
 \end{array}
\end{eqnarray*}
\caption{Differential cross sections \p{total_s} 
 as a function of the center-of-mass energy of colliding 
 partons $\sqrt{s}\leq M_D$. 
The solid and dashed lines correspond to 
 the results of the SM and the ADD model, respectively. 
The differential cross sections 
 for the like (dotted) and the unlike (dash-dotted) 
 top-antitop spin pair productions in the ADD model 
 are also depicted. }
\label{cross4}
\end{center}
\end{figure}

We are also interested in the dependence of the cross section 
 on the center-of-mass energy of parton system
 $\sqrt{s}=M_{t\bar{t}}$, 
 which is given by 
\begin{eqnarray}
 \frac{d \sigma_{tot}(pp \rightarrow t\bar{t})}{d\sqrt{s}}=
  \sum_{a,b} \int\limits_{-1}^{1} d \cos\theta
   \int\limits_{\frac{s}{E_{CMS}^2}}^1 d x_1 
   \frac{2\sqrt{s}}{x_1 E_{CMS}^2} f_a(x_1,Q^2)
   f_b\left(\frac{s}{x_1 E_{CMS}^2},Q^2\right)
   {d\sigma(t\bar{t}) \over d\cos \theta}.
   \label{total_s}
\end{eqnarray}
The results with $M_D=1$ TeV 
 are shown in Fig.~\ref{cross4}. 
The deviation of the cross section in the ADD model 
 from the one in the SM grows 
 as $s (\leq M_D)$ becomes large, 
 since the interference term 
 between the SM and the KK graviton mediated processes 
 and the squared amplitude of 
 the purely KK graviton mediated process 
 are proportional to $s^2$ and $s^4$, respectively.

\begin{figure}[H]
\begin{center}
  \epsfxsize=7.5cm
  \epsfbox{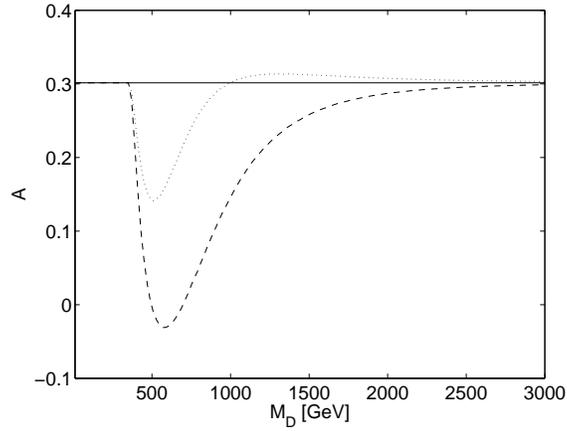}
\caption{Spin asymmetry $\cal{A}$ as a function 
 of the scale $M_D$ for the LHC with $E_{CMS}=14$ TeV. 
The horizontal line corresponds to the SM. 
The dashed and the dotted lines correspond 
 to $\lambda=1$ and $\lambda=-1$ cases
 in the ADD model, respectively.}
\label{spin0}
\end{center}
\end{figure}

\begin{figure}[H]
\begin{center}
\leavevmode
\begin{eqnarray*}
\begin{array}{ccc}
  \epsfxsize=7.5cm
  \epsfbox{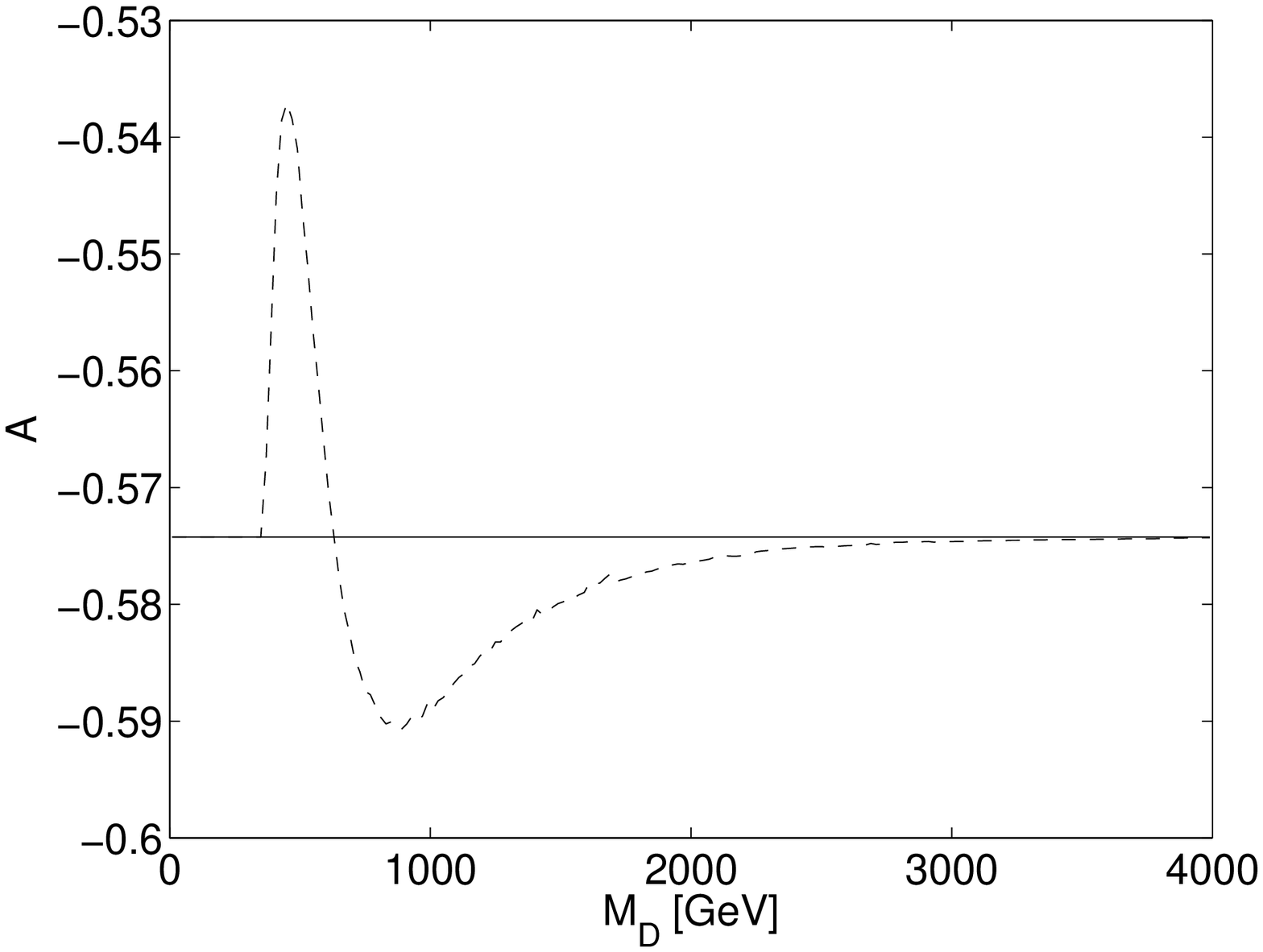} &
  \hspace{5mm}       &
  \epsfxsize=7.5cm
  \epsfbox{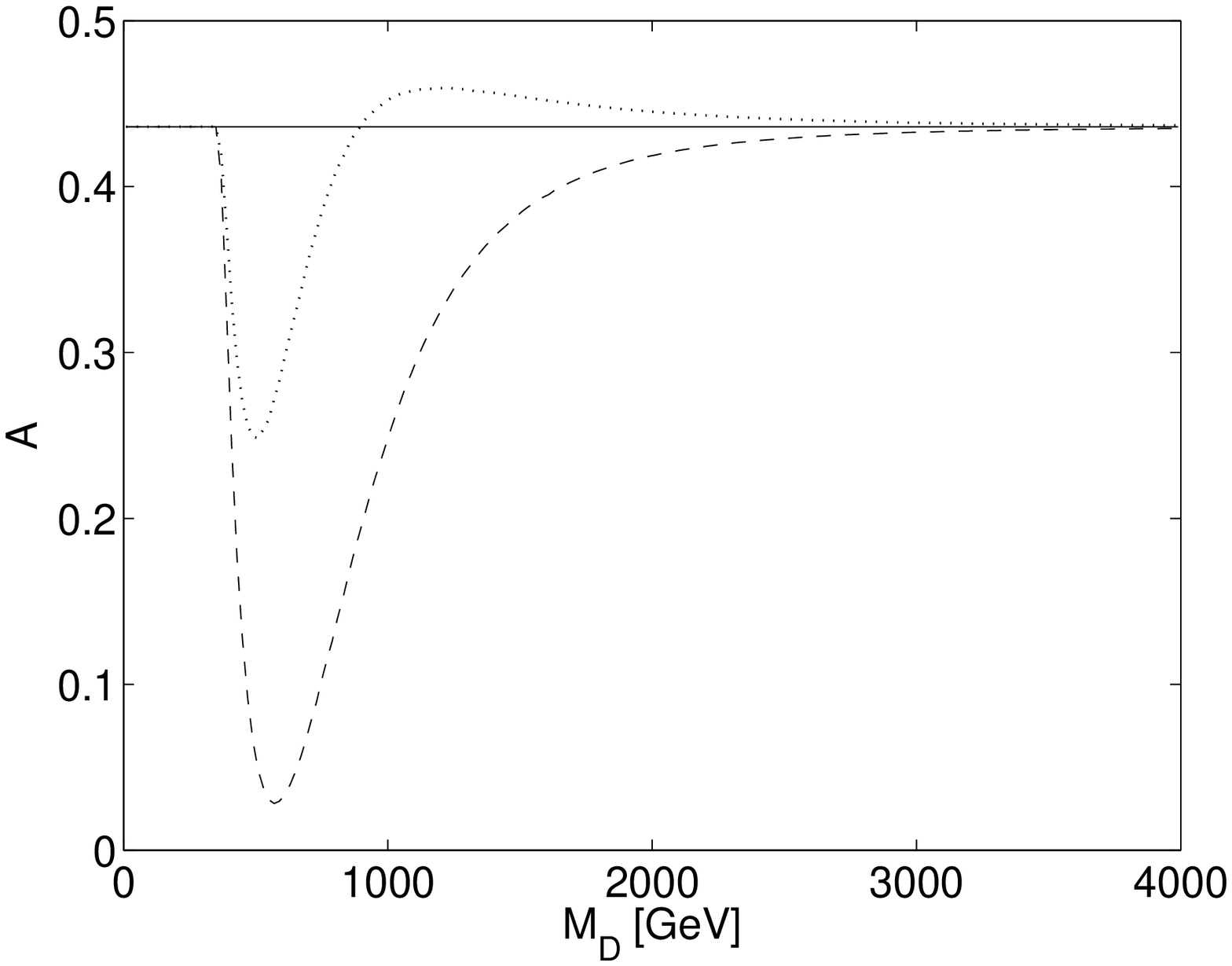} \\
\mbox{\footnotesize (a)} &
 &
\mbox{\footnotesize (b)}
\end{array}
\end{eqnarray*}
\caption{
Spin asymmetry $\cal{A}$ 
 for the $q\bar{q}\rightarrow t\bar{t}$ channel (a) 
 and the $gg\rightarrow t\bar{t}$ channel (b) 
 as a function of the scale $M_D$ 
 at the LHC with $E_{CMS}=14$ TeV. 
The horizontal lines correspond to the SM prediction, 
 while the dashed ($\lambda=1$) lines and dotted ($\lambda=-1$) lines 
 correspond to the predictions of the ADD model.}
\label{spin0_12}
\end{center}
\end{figure}

Let us show the results for the spin asymmetry $\cal{A}$. 
In Fig.~\ref{spin0}, 
 the spin asymmetry as a function of the scale $M_D$ 
 is depicted. 
We can see sizable deviations from the SM one 
 at the scale below $\sim 2~{\rm TeV}$. 
For the $q\bar{q}\rightarrow t\bar{t}$ and
 the $gg\rightarrow t\bar{t}$ channels, 
 the spin asymmetries as a function of the scale $M_D$ 
 are depicted in Fig.~\ref{spin0_12}. 
Note that for the $q\bar{q}\rightarrow t\bar{t}$ channel 
 the cross section is independent of the sign 
 of $\lambda$ (see \p{qq1} and \p{qq2}). 
As can be seen in Fig. \ref{spin0}, the total spin asymmetries 
 is dominated by the gluon fusion since
 the gluon fusion is the dominant process for the 
 top-antitop quark pair production at the LHC.

\begin{figure}[H]
\begin{center}
  \epsfxsize=7cm
  \epsfbox{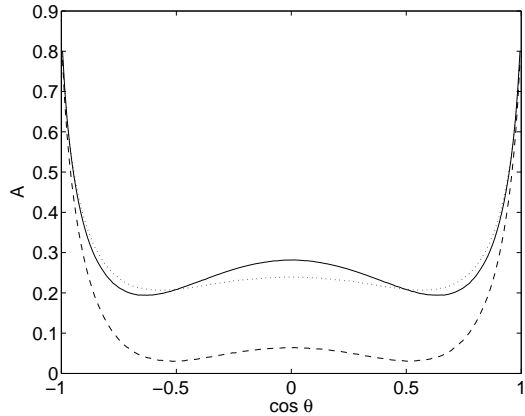}
\caption{
 Spin asymmetry $\cal{A}$ as a function 
 of the scattering angle $\cos\theta$ 
 with $M_D=1$ TeV and $E_{CMS}=14$ TeV. 
The solid line corresponds to the SM, 
 while the dashed ($\lambda=1$) line and the dotted ($\lambda=-1$) line 
 correspond to the results of the ADD model. }
\label{spin2}
\end{center}
\end{figure}

\begin{figure}[H]
\begin{center}
  \epsfxsize=7cm
  \epsfbox{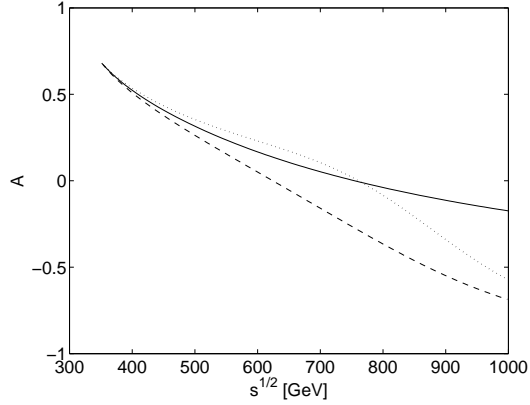}
\caption{
Spin asymmetry $\cal{A}$ as a function of 
 the center-of-mass energy of colliding partons 
 $\sqrt{s} \leq M_D$ with the scale $M_D=1$ TeV. 
The solid line corresponds to the SM, 
 while the dashed ($\lambda=1$) line
 and the dotted ($\lambda=-1$) line 
 correspond to the results of the ADD model. }
\label{spin3}
\end{center}
\end{figure}

As an analogy to Fig.~\ref{cross3} and Fig.~\ref{cross4}, 
 the spin asymmetries as functions 
 of the scattering angle (Fig.~\ref{spin2}) 
 and $\sqrt{s}$ (Fig.~\ref{spin3}) are also shown 
 for $M_D=1$ TeV. 
We can see the similar behaviors of the deviations 
 to that in Fig.~\ref{cross3} and Fig.~\ref{cross4}. 

\begin{figure}[H]
\begin{center}
\leavevmode
\begin{eqnarray*}
\begin{array}{ccccc}
  \epsfxsize=5cm
  \epsfbox{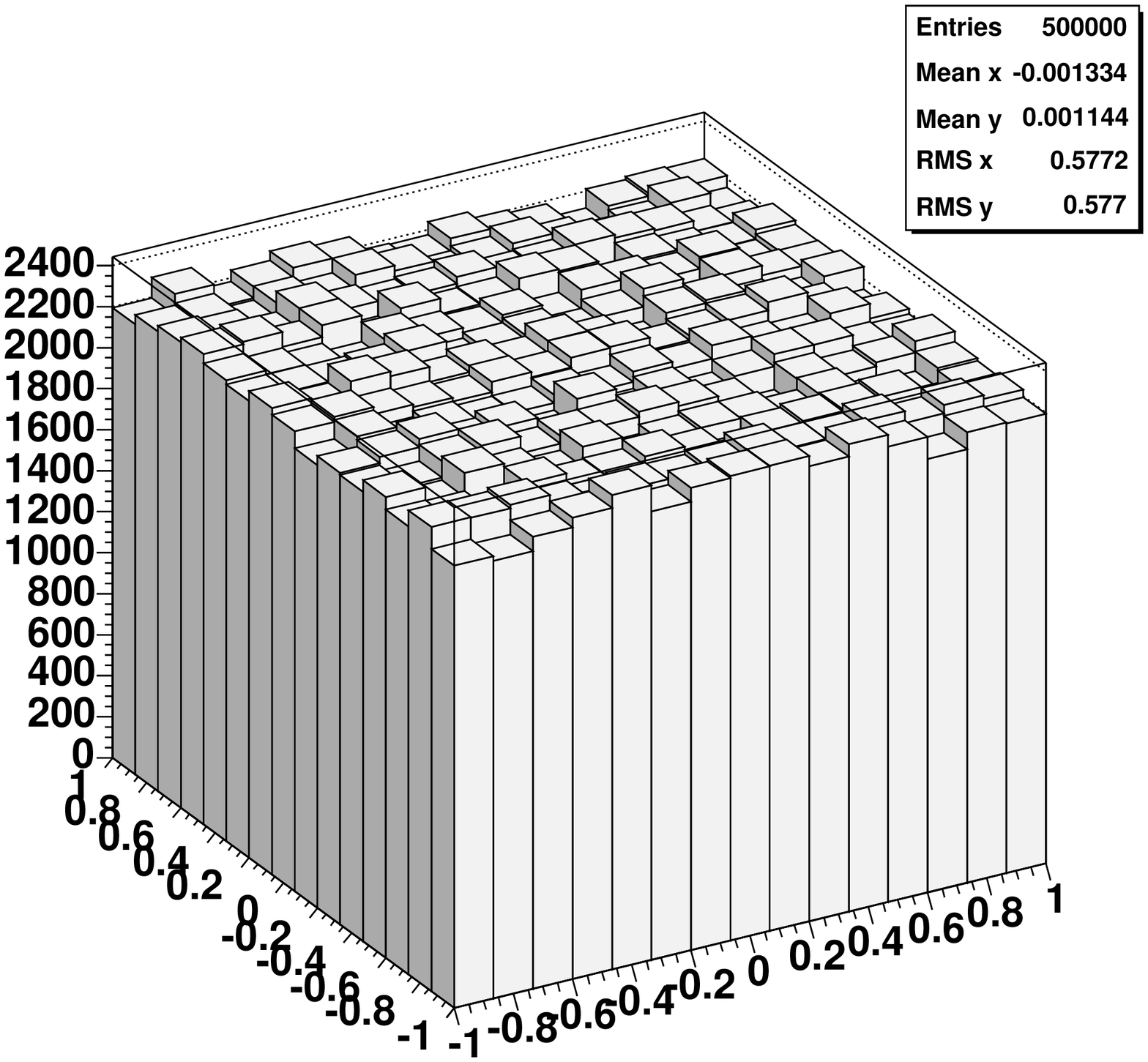} &
  \hspace{0mm}       &
  \epsfxsize=5cm
  \epsfbox{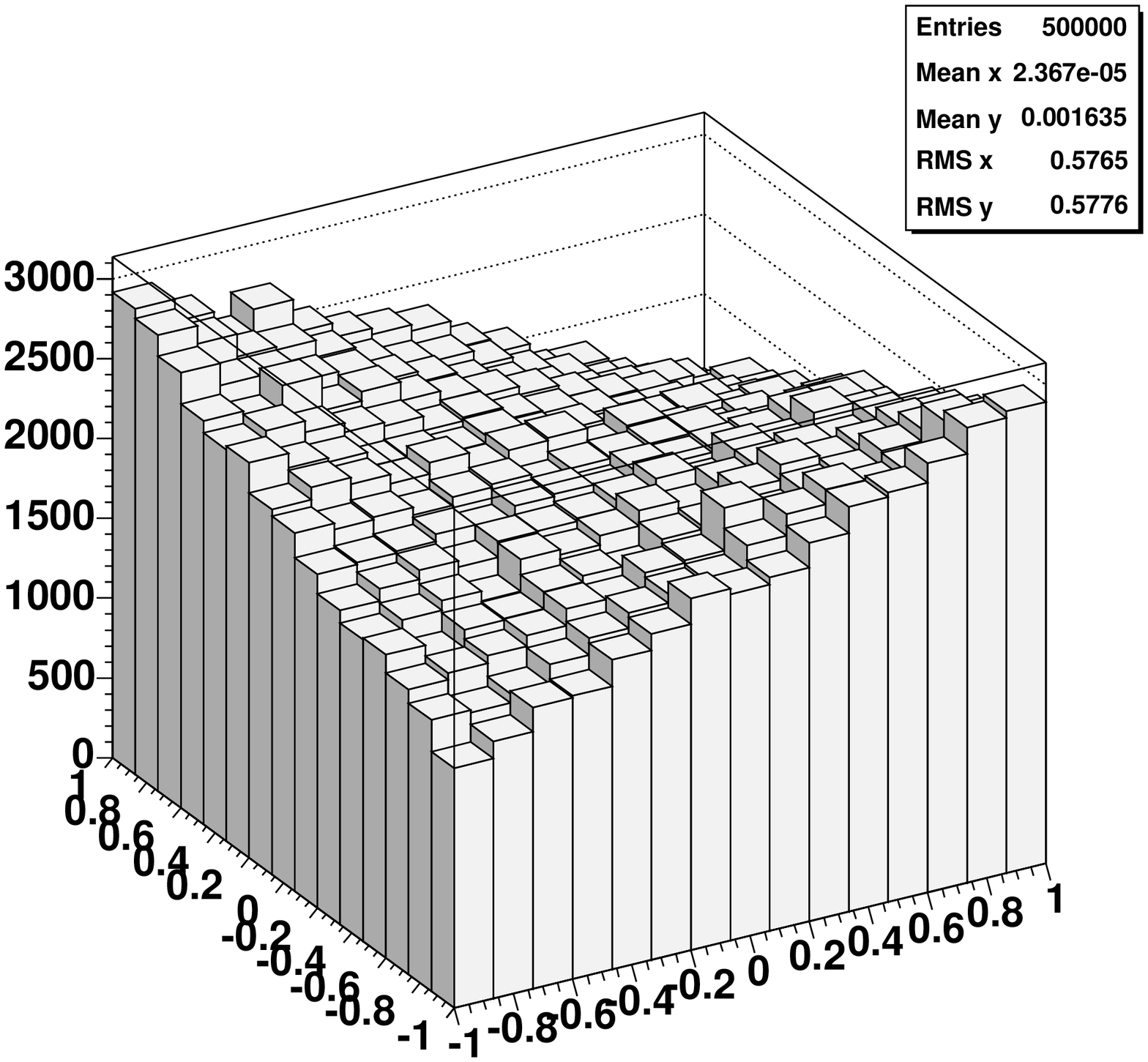} &
  \hspace{0mm}       &
  \epsfxsize=5cm
  \epsfbox{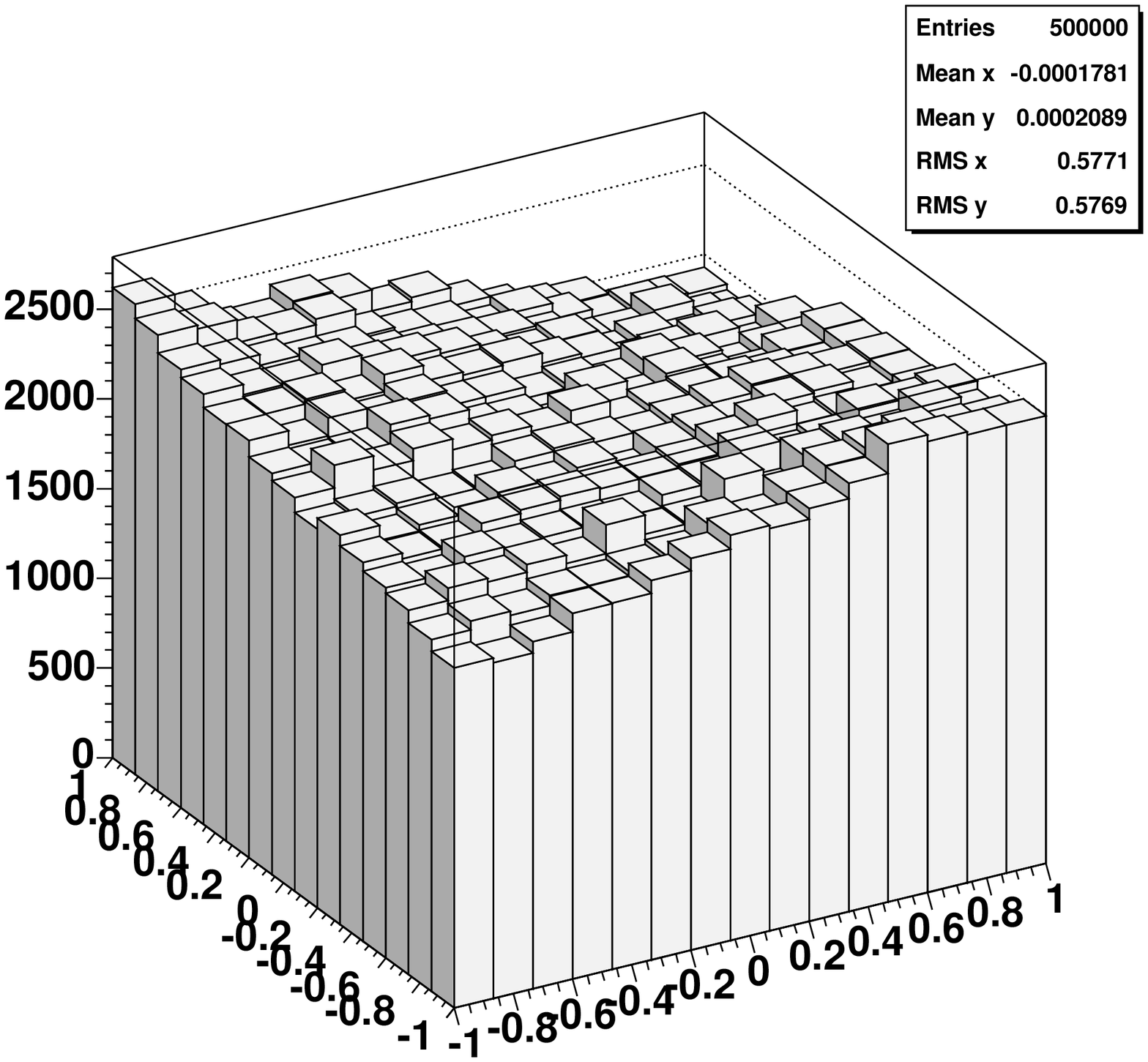} \\
\mbox{\footnotesize (a)} &
 &
\mbox{\footnotesize (b)} &
 &
\mbox{\footnotesize (c)}
\end{array}
\end{eqnarray*}
\caption{Lego plots of $\cos\theta_{l^+}$ vs. $\cos\theta_{l^-}$ 
 for the top-antitop events 
 with spin correlations ${\cal A}=0.000$ (a), 
 ${\cal A}=0.302$ (b) corresponding to the SM prediction, 
 and ${\cal A}=0.147$ (c) corresponding 
 to the ADD model prediction with $\lambda=1$, 
 $M_D=1$ TeV and $E_{CMS}=14$ TeV. 
}
\label{strechy}
\end{center}
\end{figure}

Finally, in Fig.~\ref{strechy}, 
 we show lego plots of $\cos\theta_{l^+}$ vs. $\cos\theta_{l^-}$ 
 for the top-antitop events 
 with spin correlations ${\cal A}=0.000$ (a), 
 ${\cal A}=0.302$ (b) corresponding to the SM prediction, 
 and ${\cal A}=0.147$ (c) 
 corresponding to the ADD model prediction 
 with $\lambda=1$ and $M_D=1$ TeV. 
These three plots would be distinguishable. 

%%%%%%%%%%%%%%%%%%%%%%%%%%%%%%%%%%%%%%%%%%%%%%%%%%
%
% Conclusion
%
%%%%%%%%%%%%%%%%%%%%%%%%%%%%%%%%%%%%%%%%%%%%%%%%%%
\section{Conclusion}
In theories with large extra dimensions, 
 we have studied the production of top-antitop pairs 
 and the top spin correlations at the LHC. 
There is the new contribution to 
 the top-antitop pair production process 
 mediated by the virtual KK gravitons 
 in the $s$-channel. 
We have computed the corresponding density matrices 
 and presented the various numerical results 
 based on the resultant density matrices. 
For the fundamental scale $M_D$ lower than around $2$ TeV, 
 we have found the sizeable deviations of 
 the top-antitop production rate, the spin asymmetry etc. 
 from the ones in the SM. 

%%%%%%%%%%%%%%%%%%%%%%%%%%%%%%%%%%%%%%%%%%%%%%%%%%
%
% Acknowledgement
%
%%%%%%%%%%%%%%%%%%%%%%%%%%%%%%%%%%%%%%%%%%%%%%%%%%
\vspace{1.0cm}
\noindent{\Large \bf Acknowledgements} \\

Numerical results published in this work were computed 
 with the Supercluster of the Computing and 
 Information Centre of the Czech Technical University in Prague.
N.O. would like to thank the Abdus Salam International Centre 
 for Theoretical Physics, Trieste, 
 during the completion of this work. 
The work of N.O. is supported in part 
 by the Grant-in-Aid for Scientific Research (\#15740164) 
 from the Ministry of Education, Culture, Sports, 
 Science and Technology of Japan. 

%%%%%%%%%%%%%%%%%%%%%%%%%%%%%%%%%%%%%%%%%%%%%%%%%%
%
% Appendix
%
%%%%%%%%%%%%%%%%%%%%%%%%%%%%%%%%%%%%%%%%%%%%%%%%%%
\vspace{5mm}
\noindent{\Large \bf Appendix : 
Density matrix $R_{\alpha\beta,\alpha^\prime\beta^\prime}^i$}
\vspace{4mm}

In this appendix, we give the full representation of the density 
 matrix for the top antitop pair production in the ADD model.
For the $q\bar{q}\rightarrow t\bar{t}$ process, 
 we find
\begin{eqnarray}
R_{\uparrow\uparrow,\uparrow\uparrow}^q&=&
 R_{\downarrow\downarrow,\downarrow\downarrow}^q=
 -R_{\uparrow\uparrow,\downarrow\downarrow}^q=
 -R_{\downarrow\downarrow,\uparrow\uparrow}^q \nonumber \\ 
 &=&{g_s^4 \over 9}(1-\beta^2)\sin^2\theta
  +{f_Gs^4\beta^2 \over 2}(1-\beta^2)\sin^2 2\theta, \\
R_{\uparrow\uparrow,\uparrow\downarrow}^q&=&
 R_{\uparrow\uparrow,\downarrow\uparrow}^q=
 R_{\uparrow\downarrow,\uparrow\uparrow}^q=
 R_{\downarrow\uparrow,\uparrow\uparrow}^q=  
 -R_{\uparrow\downarrow,\downarrow\downarrow}^q=
 -R_{\downarrow\uparrow,\downarrow\downarrow}^q=
 -R_{\downarrow\downarrow,\uparrow\downarrow}^q=
 -R_{\downarrow\downarrow,\downarrow\uparrow}^q \nonumber \\            
 &=&{g_s^4 \over 9}\sqrt{1-\beta^2}\sin\theta\cos\theta
 +{f_Gs^4\beta^2 \over 4}\sqrt{1-\beta^2}\sin 4\theta, \\
R_{\uparrow\downarrow,\uparrow\downarrow}^q&=&
 R_{\downarrow\uparrow,\downarrow\uparrow}^q
 ={g_s^4 \over 9}(1+\cos^2\theta)
  +{f_Gs^4\beta^2 \over 2}(\cos^22\theta+\cos^2\theta), \\
R_{\uparrow\downarrow,\downarrow\uparrow}^q&=&
 R_{\downarrow\uparrow,\uparrow\downarrow}^q
 ={g_s^4 \over 9}(-1+\cos^2\theta)
  -{f_Gs^4\beta^2 \over 2}(1+2\cos 2\theta)\sin^2\theta,
\end{eqnarray}
and, for the $gg\rightarrow t\bar{t}$ process, 
\begin{eqnarray}
R_{\uparrow\uparrow,\uparrow\uparrow}^g&=&
 R_{\downarrow\downarrow,\downarrow\downarrow}^g \nonumber \\
 &=&\left({g_s^4 \over 96}{\cal Y}(\beta,\cos\theta)
  (1+\beta^2+\beta^2\sin^4\theta)
  +{\cal Z}(\beta,\theta,s)f_Gs^2 \sin^4\theta\right)
  \beta^2(1-\beta^2), \\
R_{\uparrow\uparrow,\downarrow\downarrow}^g&=&
 R_{\downarrow\downarrow,\uparrow\uparrow}^g \nonumber \\
 &=&
  -\left({g_s^4 \over 96}{\cal Y}(\beta,\cos\theta)
   (-1+\beta^2+\beta^2\sin^4\theta)
   +{\cal Z}(\beta,\theta,s)f_Gs^2 \sin^4\theta\right)
   \beta^2(1-\beta^2), \\
R_{\uparrow\uparrow,\uparrow\downarrow}^g&=&
 R_{\uparrow\uparrow,\downarrow\uparrow}^g=
 R_{\uparrow\downarrow,\uparrow\uparrow}^g=
 R_{\downarrow\uparrow,\uparrow\uparrow}^g=  
 -R_{\uparrow\downarrow,\downarrow\downarrow}^g=
 -R_{\downarrow\uparrow,\downarrow\downarrow}^g=
 -R_{\downarrow\downarrow,\uparrow\downarrow}^g=
 -R_{\downarrow\downarrow,\downarrow\uparrow}^g \nonumber \\            
 &=& \left({g_s^4 \over 96}{\cal Y}(\beta,\cos\theta)
  \cos\theta
  +{\cal Z}(\beta,\theta,s)f_Gs^2\right)\beta^2\sqrt{1-\beta^2}\cos\theta
   \sin^3 \theta, \\
R_{\uparrow\downarrow,\uparrow\downarrow}^g&=&
 R_{\downarrow\uparrow,\downarrow\uparrow}^g
 =\left({g_s^4 \over 96}{\cal Y}(\beta,\cos\theta)
  +{\cal Z}(\beta,\theta,s)f_Gs^2 
  \right) \beta^2(1+\cos^2\theta)\sin^2\theta , \\
R_{\uparrow\downarrow,\downarrow\uparrow}^g&=&
 R_{\downarrow\uparrow,\uparrow\downarrow}^g
 =-\left({g_s^4 \over 96}{\cal Y}(\beta,\cos\theta)
  +{\cal Z}(\beta,\theta,s)f_Gs^2 \right)\beta^2\sin^4\theta,
\end{eqnarray}
where the functions ${\cal Y}(\beta,\cos\theta)$ 
 and ${\cal Z}(\beta,\theta,s)$ are defined in \p{funcs}.
%%%%%%%%%%%%%%%%%%%%%%%%%%%%%%%%%%%%%%%%%%%%%%%%%%
%
% Reference 
%
%%%%%%%%%%%%%%%%%%%%%%%%%%%%%%%%%%%%%%%%%%%%%%%%%%

\newpage

\vspace{-5mm}

%%%%%%%%%%%%%%%%%%%%%%%%%%%%%%%%%%%%%%%%
%
% Spin Asymmetry
%
%%%%%%%%%%%%%%%%%%%%%%%%%%%%%%%%%%%%%%%%


\end{document}